\documentstyle[psfig]{mn}
\def\ha{$H\alpha$}
\def\hb{$H\beta$}
\def\la{$Ly\alpha$}

\title{The Nature of the Broad--Line--Region in the Radio--Loud AGN 3C390.3}

\author[W.Wamsteker et al.]{Willem Wamsteker$^{1,2}$,
        Wang Ting-gui$^{3}$, Norbert Schartel$^{1,2}$, 
        Roberto Vio$^{4}$ \\
      $^{1}$ESA IUE Observatory, P.O. BOX 50727, 28080 Madrid, Spain \\
      $^{2}$Affiliated with the Astrophysics Division, Space Sciences Department,
            ESTEC, the Netherlands. \\
      $^{3}$Center for Astrophysics, University of Science and Technology of
            China, Hefei, 230026, Anhui, P.R. China \\
      $^{4}$Astronomy Department, University of Padova, Vicolo dell Osservatorio
            5, 35122, Padova, Italy. }

\date{Accepted 1997 February 5, 
      Received 1996 January 2; 
      in original form 1996 January 2}
\pubyear{1997}

\begin{document}

\maketitle

\begin{abstract}
We present the results of an analysis of the ultraviolet and X--Ray variability of 
the 
Broad--Line--Radio Galaxy 3C390.3 over 15 years. The UV continuum of 
3C390.3 showed large variations with amplitudes of up to a factor of 10. We find: 

\noindent
(1) The variations of  CIV and \la~ are highly correlated with the UV continuum, 
and are delayed with respect to the  continuum variations by  50--110  days with 
the red wing of both CIV and \la , leading the blue wing;

\noindent
(2) The CIV/\la~ ratio is positively correlated with both the  continuum flux and 
UV line strength, a behavior different from other AGNs studied so far; 

\noindent
(3) The blue sides of the \la~ and CIV profiles are similar to the blue side of the 
Balmer lines, while the  red sides are different, suggesting a different origin for 
the red peak in the Balmer lines. 

The  X--Ray spectra of 3C390.3 observed with ROSAT can be well fitted by a 
single power-law at Galactic absorption with a spectral slope of  $\alpha$ = 0.9. 
The overall optical, UV to X--Ray spectrum can  also be described by a single 
power law with $\alpha_{uvx}$ = 0.89, indicating a very weak or no big blue 
bump. The unusual behavior of CIV/\la~ variations might be related to this hard 
ionizing continuum in 3C390.3. 

Our results suggest: (1) The broad CIV and \la~ emitting gas is infalling towards 
the central object in 3C390.3; (2) The overall behaviour of the  CIV/\la~ ratio and 
the absence of a big blue bump, strongly indicate the coexistence of optically 
thick as well as optically thin BLR clouds; (3) Assuming circular symmetry and 
predominantly circular motion, the BLR gas is situated at 83$\pm$25 lightdays 
from the central source; (4) Under these assumptions and with the derived circular 
velocity of $v_{rot} \simeq$ 2850 km~s$^{-1}$, the central mass inside this 
radius is confined to  $1.3~10^8 M\odot < M_{CM}< 4.0~10^8 M\odot$; (6) 
Comparing our results with those obtained from VLBI and observations of the Fe 
$K\alpha$ line, suggests the association of the BLR with a disk, inclined at 
98$\pm12$~degrees with respect to the direction of superluminal motion of the 
radio blobs.
\end{abstract}


\section{Introduction}

Monitoring emission line and continuum variability has proven to be a powerful 
method to probe the distribution, as well as the physical and kinematic state of the 
line-emitting gas surrounding Active Galactic Nuclei (see Peterson, 1993 for a 
review). The photo-ionized gas responds to the energy input from the continuum 
source with a delay indicative of the light crossing time for the Broad Line Region 
(BLR). Campaigns by the AGN Watch collaboration, with a high and even 
temporal  
sampling, have been carried out for a few bright Seyfert galaxies, NGC5548 
(Clavel et 
al., 1991; Peterson et al., 1991, 1992; Stirpe et al., 1993; Korista et al., 1995), 
NGC4151 
(Clavel et al.,1991a; Crenshaw et al.,1994), NGC3783 (Reichert et al.,1994) and 
Fairall-9 (Rodriguez--Pascual et al., 1997). These campaigns have clearly 
demonstrated the power of this method. The results are important with respect to 
the 
following questions on the nature of the BLR: (1) Does BLR size scale as 
r~$\propto$~L$^{1/2}$?  (2) How does the emission line ratio vary with 
continuum? 
(3) What is the kinematics of BLR? The answer to the first question seems yes. 
Up to 
now, all results for Seyfert galaxies and quasars are consistent with 
r~$\propto$L$^{1/2}$ (Peterson,1993; Kaspi,1994). Concerning the second 
question, 
the CIV/\la~ ratio  decreased with increasing continuum luminosity in NGC5548, 
Fairall 9, 3C120 ( Clavel, Wamsteker and Glass,1987; Wamsteker et al.,1990; 
Gondhalekar, 1992). Three possibilities have been proposed to explain this 
behavior: 
a change in the shape of the ionizing continuum (Clavel et al.,1988), optical thin 
BLR 
material (Wamsteker \& Colina,1986) or a mixed population of optically thin and 
thick BLR clouds (Shields et al.,1995). Regarding the kinetics of the BLR, there 
exists 
some evidence for infall in Fairall 9 (Recondo et al.,1997; Koratkar \& 
Gaskell,1991). 
Accurate determinations of the size of the broad emission line region and, in some 
cases,  
the responsive--average geometry (Transfer Function, TF) have been derived from 
these data (Krolik et al.,1991; Horne et al.,1991; Ferland et al.,1992). While 
detailed 
interpretation of  the Transfer Function is subtle and to a significant extent model 
dependent (Maoz,1994), these analyses show: (1) the broad emission lines 
respond to 
the continuum variations very fast and show a dependence on the ionization level 
of 
the ions, with the shortest response time for the highest ionization lines (Korista et 
al.,1995; Peterson,1993), and  (2) the line-emitting gas is not dominated by radial 
motion. 
The red wing of CIV has been suggested to lead to the blue wing in NGC5548 
(Korista et 
al.,1995; Wanders et al.,1995), but this result may be a consequence of the 
blending of 
CIV with the  very broad wings of HeII. However, the general validity of the 
above 
results remains still an open question. There are still too few objects for which 
BLR 
sizes are 'reasonably' determined and furthermore, all these analyses are limited to 
the 
radio quiet objects with big blue bumps and regular emission line profiles. 
Although observations with high temporal sampling  are only available for these 
four 
Seyfert I galaxies, one can still use the long term, unevenly sampled, observations 
with the International Ultraviolet Explorer (IUE) and ground based optical 
observations to derive estimates of the size and kinematics for other AGNs. 
Especially for radio galaxies and other galaxies without a big blue bump, where 
no systematic monitoring has been done, this could be of interest. In this paper, 
we will present an analysis of the variability of 3C390.3, based on the UV and X--
Ray observations made between 1978 and 1992.

3C390.3 (1845+796) is a FRII radio galaxy at redshift z=0.0561. Superluminal 
motions were observed at radio wavelengths (Alef et al.,1988). Large amplitude 
variations have been reported in optical, UV and X--Ray bands (Lloyd,1984; 
Veilleux\& Zheng,1991; Clavel \& Wamsteker,1987; Inda et al.,1994). The IUE 
observations of this object prior to 1986 have been discussed by Clavel \& 
Wamsteker 
(1987) and  variations of the narrow \la~ and CIV  were reported. Also, line 
variations 
confined to a restricted velocity range in the broad UV emission lines have been 
reported by Wamsteker \& Clavel (1989).  Zheng (1996) discussed independently 
the UV variability based on the same observational material as used here. The 
profiles of the Balmer lines are very broad 
and double peaked (Osterbrock, Koski and Phillips,1977). Veilleux \& Zheng 
(1991) 
found that the peak at the blue side of \hb~followed the variation of the continuum 
closely and drifted in velocity with time, but the peak at the red side (v = 3000 
km~s$^{-1}$) did not. 
The double peaked profile and its variations were interpreted to indicate the 
presence 
of a relativistic accretion disk (P\'erez et al.,1988; Chen \& Halpern,1989) or to 
suggest  biconical outflow (Zheng, Veilleux \& Grandi,1991). From observations 
with the Ginga satellite a weak K$\alpha$~ 
emission (6.4 keV) of Fe was found in the X--Ray spectrum of 3C390.3 by Inda 
et al. (1994) and new observations with ASCA have resolved the line and are 
discussed by Eracleous et al. (1996).  The infrared to optical spectrum is very 
steep, and no big blue bump 
appears to be present in this object (Miley et al.,1984). In \S 2 of this paper, we 
describe the data and the processing methods. The results are presented in \S 3. In 
\S 
4, we discuss the implications of these results. Finally, the conclusions are 
summarized in \S 5. Zheng (1996) discussed the same data as we present, the 
differences between his results and ours are mostly due to the different methods 
used 
for the determination of the UV continuum shape and the application of a line 
model derived from the optical data to the UV lines. We show below that the 
application of such model to the \la~ and the CIV line is not justified. 

\section{DATA REDUCTION AND ANALYSIS}

A total of 39 short wavelength (SWP: 1150--1980 \AA) IUE observations have 
been 
taken between 1978 and 1992 (see also Courvoisier and Paltini,1992). The 
observation dates and continuum fluxes are listed in Table 1.  We used the 
geometrically and photometrically corrected Extended Line--by--Line (ELBL) 
files, 
produced by the IUESIPS reduction, which were cleaned of cosmic-ray hits and 
other 
blemishes. The UV spectra were then extracted from the ELBL files by a modified 
Gaussian extraction (GEX) technique (Reichert et al.,1994) using the routines 
written 
at the IUE Data Analysis Center (IUEDAC) and calibrated using the standard IUE 
sensitivity curve, including the corrections for time and temperature (THDA) 
sensitivity degradation of the SWP camera. Galactic reddening of E(B-V)=0.085 
has been applied to all 
spectra using the reddening curve of Seaton (1979). To bring all spectra on the 
same wavelength scale (i.e. correct for centering errors in the IUE Large 
Aperture), the narrow CIV lines were fitted by a single Gaussian and the 
central wavelengths were made to coincide at  1635.9 \AA.

The continuum measurements in Table 1 are made in two pseudo line--free bands 
at 
1330~$-$~1370\AA~  and 1775~$-$~1825\AA~  in the rest frame of 3C390.3. 
The 
continuum flux is the mean over the bandpass, and the error is the standard 
deviation 
of the mean over this bandpass. The continuum at 1800\AA~ is contaminated by 
the 
UV FeII blends lines at the 10\% level, as estimated from the mean spectrum (see 
also 
below). The continuum level underneath the emission lines was determined by a 
power-law derived from these two continuum windows.

 Veilleux et al. (1990) found that the blue peak of \hb~ extends over the velocity 
interval of (-7000, 0) km~s$^{-1}$ relative to systemic velocity. This velocity 
range 
corresponds to that in which most of the blue part of CIV is emitted. To maintain 
the differences reported previously for different velocity domains of the broad 
lines (Wamsteker \& Clavel,1989),   we subdivided both CIV and 
\la~ in three parts: the blue-side (-7000 to -1800 km~s$^{-1}$); the 
red-side (1800 to 7000 km~s$^{-1}$) and the core (-1800 to +1800 km~s$^{-
1}$). 
The red side was chosen in symmetry with the blue side. More complex fitting 
techniques, as e.g. applied by Zheng (1996), are not necessarily superior with S/N  
levels as in the IUE spectra of 3C390.3. Also, the detailed application of the line 
model derived from the optical data is not valid in this case (see also Figure 5).

Although the overall lightcurves of Zheng (1996) are quite similar to those 
presented here, we have, in view of the rather different analysis techniques,  not 
made a further detailed comparison with his  data and the data in Tables 1 and 2. 
(N.B. the data in our Table 1 and 2 have been corrected for reddening as above). A 
comparison of CIV and \hb~ profile (see Figure 5) shows: (1) The blue side of 
CIV and Balmer line profile is very similar: the peak velocities are the same 
within the uncertainty; (2) The red sides of profiles are different, Balmer line 
shows a strong red bump peaked around +3000 km~s$^{-1}$, however, only a 
weak second peaked around +6000 km~s$^{-1}$~ is visible in CIV; (3) In the 
\la~ profile, a strong core component, which is much less pronounced in the 
Balmer lines, is present. The variation on the blue side of CIV and \hb~ is similar 
in both lines. This suggest that the blue sides in the different lines have the same 
origin, while the prominent red peak seen only  in the Balmer lines, is of a 
different origin. The absence of any correlation of the red peak of  \hb~ with the 
continuum  (Veilleux \& Zheng,1991) is in agreement with the notion that this 
part of the line is not associated with the gas seen in the UV lines and does 
not respond rapidly to chances in the continuum intensity.

The emission line flux above the continuum was  integrated over the 
corresponding wavelength intervals. The core region in this case is of course 
strongly contaminated by the important contribution of the narrow line, while the 
red side of \la~ is contaminated somewhat by the weak NV $\lambda$1240 line. 
However, the CIV red and blue side are almost free of contamination from any 
other lines. The total flux of CIV and \la~ was derived by integration over the full 
velocity interval from -12,000 to +12,000 km~s$^{-1}$. 
The line fluxes measured in this way for the emission lines are given in Table 2.

From 1978 to 1993 pointed  X--Ray observations of 3C390.3 have been made at 
17 
epochs with different instruments. The observations from HEAO1-A2, Einstein 
IPC 
and MPC are  from Malaguti, Bassani \& Caroli (1994), while the EXOSAT, and 
Ginga results are from Inda et al. (1994). The photon spectra were fitted by a 
single 
power-law of the form: n$_{ph}$(E)~=~A~E$^{-\Gamma}\;\exp(-
\sigma_EN_H)$, 
where E, $\sigma_E$ and $N_H$ are the photon energy, photo-absorption cross 
sections and the absorption column density, respectively, and  $\Gamma$ is the 
photon-index. The 4 ROSAT observations taken between 1991 and 1993 were 
retrieved from the ROSAT database at Max--Planck--Institut f\"ur 
Extraterrestrische 
Physik and processed independently  with EXSAS (Zimmerman et al.,1994). The 
source counts were extracted from a 2' circle centered on the source centroid. The 
background is  determined from the counts within an annulus of inner radius of  2'  
and outer radius of 5'. All sources detected with a likelihood greater than 10, 
defined 
as the area of 2' radius around the center of light, are excluded from the 
background, 
and the exposure time was corrected for the dead time. The gain calibrated source 
and 
background counts are  corrected individually for vignetting and finally binned in 
pulse  height spectra. The pulse  height spectrum is  modeled with a single 
power--law, folded with Galactic absorption and  the detector response function of 
ROSAT. 
The 
absorption cross sections are from  Morrison \& McCammon (1982). In most 
cases, a 
single power-law model fits the spectrum very well (see Table 3). More 
complicated 
models, such as a power--law and black-body components, do not improve the 
quality 
of the fit and are not justified by the signal--to--noise in the data. The results for 
all X--ray data are given in Table 3. The results presented here are similar to the 
high resolution spectral data presented by Eracleous et al. (1996).

The optical spectra used for 3C390.3 were taken at Lick Observatory and 
calibrated 
using [OIII] line flux (refer to \$2 of Veilleux \& Zheng, 1991 for details).

\section{3. RESULTS}

\subsection{ Variability Pattern}

Figure 1 shows the light-curves for the X--rays, the UV continuum at 1800 
\AA~and 
the CIV line (the lightcurves for \la~ are very similar in shape to those of CIV, 
although 
with smaller amplitude; see Table 4). The UV continuum light curve shows a few 
large events which typically last one to two years and have amplitudes of up to a 
factor of 5. They also suggest the presence of smaller and faster events which rise 
and 
drop in brightness with a factor of 2 in tens of days. Because of the sparse 
sampling, 
most small events can not be resolved and will not be discussed in what follows. 
The 
UV continuum reached a maximum state in late 1991 with a subsequent 
brightness 
decrease after that. The emission line light curves (see Figure 1) are similar to that 
of the continuum for the 
big events, except for the apparent presence of a delay in the peak brightness of 
some 
100 days. After 1988 (JD 2,447,600) the core and red light-curves of CIV and 
\la~ show two well-defined events, while in the blue part of the line an additional 
early 
peak appears present. This ill-defined first peak in the blue part of the lines is 
possibly 
due to the longer response time of blue part (see below \S 3.3) to a bright state in 
1988 
which was not observed by IUE. The Ginga observation of November 1988, 
which is 
the highest X--Ray flux observed between 1978 and 1993 (see Figure 1), suggests 
that 
an additional strong continuum event may have occurred in the latter part of 1988. 

During the IUE observations, the UV continuum flux varied by a factor larger 
than 10 
although the actual dynamic range (R$_{max}$) of the continuum variations is 
difficult to determine due to the fact that the lowest flux, in 1984, is defined by a 
relatively noisy spectrum (SWP22624). The variability parameters 
(F$_{var}=\sigma/\bar{F}$) and $R_{max}$~ for line and continuum are given 
in 
Table 4. Since $R_{max}$ is rather uncertain because of the large 
error associated with the lowest brightness state as well as the sparse temporal 
coverage 
in the sampling, we give in Table 4 also $R_{max}$~ ($JD>7600$), which only 
considers the data taken after 1988. This should give a better evaluation of the 
variability, since this part of the lightcurve is adequately sampled with respect to 
the variability time scale. The measurement errors for constant flux (non-variable) 
source are 
assumed to be normally distributed and we estimated the error in 
$F_{var}$~ through a Monte-Carlo calculation for data sets which are consistent 
with the measurements. 
It is clear from Table 4 that the UV continuum shows largest variability 
parameter, 
followed by CIV and \la, but  note that $R_{max}$ for CIV is nearly as 
large as that for the continuum at 1350 \AA. The variability parameters for the 
blue and red-side of the lines are the same within the errors. The core is much less 
variable than the blue and red side, due to contamination by the narrow 
line. Correcting for the narrow--line contribution, derived from the regression of 
the core flux versus continuum relation, both F$_{var}$~ and $R_{max}$ for the 
broad line flux in the core, are found to be equal to that of 
the red and blue sides of the line. Even though the variability parameters at 1350 
\AA~ and at 1800 \AA~ are the same within the errors, the short wavelength 
continuum seems to be more variable than that at 1800 \AA. 

The mean ultraviolet spectrum is shown in the upper panel of Figure 2, the 
variance is plotted in the lower panel, and the middle panel shows the normalized 
variance spectrum (division between the variance spectrum shown in the bottom 
panel and the average spectrum in the upper panel). The variance spectrum 
highlights the strong blue asymmetry in the lines also visible in the average 
spectrum. On the other hand, the flat bottomed depressions in the normalized 
variance spectrum at the positions of the broad lines indicate that the variation 
amplitude over the broad lines is essential the same over the full velocity width of 
the broad lines. The narrow depressions at the central positions correspond to the, 
much less variable, narrow lines. The many features between 1820 and 1950\AA~ 
(observed wavelength) present in the average spectrum, but absent in the variance 
spectrum, are most likely associated with the UV FeII blends and contribute at 
least 10\% of the flux in the 1800\AA~ bandpass. 

\subsection{Time Series Analysis}

Cross correlation methods have been used to determine if the variations are 
correlated and if there is a time lag between continuum and line lightcurves. Two 
different correlation functions were calculated for the UV light curves of 3C390.3: 
the 
interpolated cross correlation function (CCF, Gaskell \& Peterson,1987) and the 
discrete cross correlation function (DCF, cf. Edelson \& Krolik,1988). We have 
discarded the data pairs which required interpolated data points with separations 
larger 
than 200 days for the calculation of the CCF. The CCF are normalized using only 
the 
data points which are used for calculation of CCF at a particular lag (White \& 
Peterson,1994). The DCF's were calculated as described by White \& Peterson 
(1994).

Firstly, the continuum light--curve at 1800 \AA~ is cross correlated with itself to 
generate the continuum auto--correlation function (ACF). The continuum light 
curves 
are then cross correlated with  each line feature light curve to get the 
corresponding 
CCF and DCF. Also, the light curves of the different parts of CIV and \la~ are 
cross--correlated with each other. To calculate the DCF's, the sampling interval 
was chosen to be 50 days, which is slightly more than half  the average interval of 
the observations. Figure 3 shows the ACF, as well as the CCF's and DCF's for 
CIV, with error bars (since the lightcurves for \la~are very similar to those of CIV, 
the correlation functions are too). The ACF of  window sampling is overplotted 
(refer to Gaskell \& Peterson,1987) on the continuum ACF. To the first order, the 
CCF's and DCF's are  in all cases similar, in spite of the slight differences to be 
expected for such an unevenly sampled time series (White \& Peterson,1994). 
Table 5 gives the main characteristics of CCF's. The parameter $\Delta$t$_{peak}$ 
gives the time lag (in days) for r$_{max}$, the peak value of CCF. The centroid 
measures at 0.5 r$_{max}$ give similar results. The errors were calculated with a 
Monte Carlo (MC) calculation, assuming a Gaussian error distribution and represent the 68\% 
range. The MC calculation was done over a range of $\pm$~ 500 days to avoid 
the overlap of the two indivdual peaks in  the lightcurves at JD7800 and JD8500. 
This is only significant for the CCF of the core lightcurves, where the amplitude 
of the emission line lightcurve is signifcantly reduced by the contribution of the 
narrow line. A positive value for $\Delta$t$_{peak}$ indicates that the variation 
of the second quantity lags that of the quantity in the first series. 

The FWHM of continuum ACF is about 200 days, although larger than the 
average 
observation interval (90 days), the sampling is only just below the Nyquist 
frequency 
for the long term variability mode. The fact that the FWHM of the ACF for 
CIV(total)  
is 400 days, twice the width of continuum ACF, implies that the continuum 
variation 
is faster than the light--crossing time of the BLR. This is in good agreement with 
that 
inferred from direct inspection of light-curves in Figure 1, especially for the large 
and 
reasonably sampled events after 1988. It is clear that emission line and continuum 
variation are highly correlated, with r$_{max}$ in the range from 0.7 to 0.9, and 
the 
variations of CIV and  \la~ (not shown in Figure 3) lag those of the continuum by 
about 100 days. Taking the average value of  $\Delta$t$_{peak}$ for both CIV 
and \la~ from Table 5, we find for the red part of the line $\Delta$t$_{red}$ = 
57$\pm$18; for the core $\Delta$t$_{core}$ = 80$\pm$41 and for the blue side 
$\Delta$t$_{blue}$ = 110$\pm$17 (average errors, see Table 5). 
It is thus clear that the red side of both  CIV and  \la~ responds significantly 
earlier to the continuum variation than the blue side. Direct cross correlaton of the 
blue and red sides of the lines, confirms a time delay between the two of 
$\Delta$t$_{R-B}$ = 46$\pm$17 days,consistent with the results from the line 
versus continuum cross correlation. The extra peak in the blue side lightcurve 
after 1988 (see also \S 3.1) can now be understood in the context of the 
differences in the delay between the red and blue sides of the emission lines. The 
Ginga observation of November 1988 --the highest X--Ray flux observed between 
1978 and 1993 (see Figure 1)-- suggests the existence of a large continuum event 
in the latter part of 1988. Since the first UV observation is 130 days after the X--
ray peak, this supports the idea that the line response to the 1988 event had 
already been completed for the red and central parts of the line at the start of the 
UV observations in 1989, while only the tail end of the response of the blue side 
of the line was still observed.  

\subsection{Variation of Emission Line Ratio}

  The CIV/\la~ line ratio is positively correlated with the continuum flux but 
shows considerable scatter and is shown in Figure 4(right). This behaviour 
contrasts with that shown by the other AGN studied in detail where the values of 
$R_{max}$(CIV/\la) for NGC 5548 and NGC 3783 are found to be 0.95 and 0.61, 
respectively as compared to a factor of three in 3C390.3. Also, the CIV/\la~ ratio 
tends to decrease slightly with increasing continuum brightness, contrary to the 
behaviour shown in Figure 4 (right). From Figure 2 (middle) and Table 4, it is 
clear that this relation is in part driven by the larger variation in the CIV line as 
compared to \la~ (by a factor of two). Regression analysis for the relation between 
\la~ and CIV for all components shows that a linear relation between the two lines 
does not give a statistically significant description of their relation 
($\chi^2$~$_{red}$~$>1.6$).

The correlation with the continnum is better for the red side, for which the 
Spearmann rank correlation 
coefficient (r$_s$; Press et al.,1986) is 0.67 (corresponding to the null correlation 
probability Pr = 0.3$\times$10$^{-5}$), than for the blue side (r$_s$=0.58, Pr = 
10$^{-4}$). The data after 1988, with a  sampling frequency of 80 days for a 
doubling time of the brightness variations of 250 days can be considered to be 
sufficiently sampled to describe the variation of the two large events. For these 
data the correlation disappears for the blue side ($r_s$= 0.207, Pr= 0.44) but still 
persists for the red component ($r_s$ = 0.775, Pr = 0.4$\times$10$^{-3}$). This 
lack of correlation is clearly related to the dissimilar light-curves for the blue side 
and continuum, and due to the larger delay found for the blue part of the line: the 
blue sides of both \la~ and CIV are poorly correlated with $F_{1800}$ with 
r$_s$=0.448 (Pr = 0.08) and r$_s$=0.401 ( Pr = 0.23 ) for the data after 1988, but 
the red side is well correlated with continuum with r$_s$=0.623 (Pr = 0.01) and 
r$_s$=0.828 (Pr = 10$^{-4}$), respectively. To overcome the problem due to the 
delays in line response for the emission lines one could resort to correcting the 
correlation diagram in Figure 4 (right) for the delay determined above, however 
the sampling of the lightcurves is not sufficient to apply such correction, as is 
clearly illustrated by the X--Ray observed event in 1988. An alternative approach 
to correct the 
relations for time delay effects, is to use the fact that the CIV line is well 
correlated with the continuum at similar amplitude, and use the CIV line intensity 
as a measure of the continuum. Figure 4 (left) shows the CIV/\la~ ratio  for both 
the red side and the blue side as a function of the blue  and red I(CIV), 
respectively. The relation in Figure 4 (left) is very well defined, in contrast to 
CIV/\la~ relation with the continuum, shown in Figure 4 (right), and the CIV/\la~ 
ratio for the two sides of the lines merge indistinguishable into each other even 
though they cover different domains in the diagram (red: $0<I(CIV)<80$ and blue 
: $9<I(CIV)<120$ , units: ~10$^{40}$ erg~s$^{-1}$~cm$^{-2}$). This relation 
shows a tight correlation for the lower intensity levels ($I(CIV)<60$ 
corresponding to  $F_{1800}$ = 0.6~10$^{40}$ erg~s$^{-1}$~cm$^{-
2}$~\AA$^{-1}$), while at higher intensities the CIV/\la~ ratio flattens out and 
appears to become constant.

\subsection{Ionizing Continuum Shape}

The big blue bump, often seen in the Seyfert galaxies and QSOs, is very weak or 
absent in 3C390.3. The mean ratio of 
F$_\lambda$(1350\AA)/F$_\lambda$(5125\AA) for the three nearly simultaneous 
(within 15 days) observations  in optical (19 May 1980, 28 May 1984 and 26 
August 1987) and UV (26 May 1980, 5 June 1984 and 12 August 1987)  is 4.36 
(after correction for the Galactic reddening), corresponding to an optical to UV 
spectral index (F$_\nu\propto \nu^{-\alpha}$ ) $\alpha_{optUV}$ = 0.9. The 
contamination by stellar light in optical is small, because the MgI and FeI 
absorption features at 5170--5210 are very weak (Veilleux \& Zheng,1991). The 
mean F$_\nu$(1keV)/F$_\nu$(1350\AA) for 4 UV and X--Ray observations with 
separation of less than 20 days, is 0.03, corresponding an UV to X--Ray spectral 
index $\alpha_{uvx}$ = 0.89. 

Although Kruper et al. (1990) claimed the existence of a strong soft X--Ray 
excess in the Einstein IPC spectrum with a very large absorption, no soft excess is 
suggested in any of four ROSAT X--Ray observations. The ROSAT spectra can 
be well modeled by a single power law with a photon spectral slope index in the 
range between 1.86 to 2.20, very similar to the results of Eracleous et al. (1996).
The soft X--ray flux decreased by a factor 2.5 between 1992 and 1993, and the 
soft X--Ray spectral slope is consistent with a constant value $\alpha$= 0.92 
$\pm0.9$~, which is slightly steeper than the spectral index from the Ginga data 
in the ME band of 0.77 and 0.54 for 1988 and 1991 respectively (Inda et al.,1994). 
The steeper soft X--ray spectral slope for the observation of 1991 is certainly due 
to the large fitted $N_H$~(Table 4). The spectral indices of optical to UV, UV to 
X--ray, soft X--ray, and ME X--Ray are all very similar for 3C390.3, strongly 
suggesting that the big blue bump is very weak or absent in this object. The 
simultaneous observations of 3C390.3 made during the RIASS campaign (de 
Martino et al.,1992) discussed by Walter et al. (1995) confirm the weakness of the 
big blue bump in 3C390.3. 

It is interesting to note that the spectral energy distribution and the spectra 
variabililty character of 3C390.3 from the optical to the X--Rays is quite similar to 
those found, from extended multi--frequency observations, for the BL Lac object 
PKS2155-304 (Edelson et al., 1995). The only distinguishing part  is the fact that 
the turnover seen in this BL Lac object at higher energies, is not seen in 3C390.3. 

\section{DISCUSSION}

We will here discuss the implications of the results described above in the context 
of 
the three questions posed in section~1.

\subsection{Does the Size BLR scale as L$^{1/2}$ ?}

If the BLR in different objects can be characterized by the same physical 
parameters (ionization parameter U and particle density $n_H$) and the ionizing 
continuum shape does not vary much from object to object, the size of BLR scales 
with continuum luminosity through, 
\begin{equation}
   r\;=\;(\frac{N_{ion}}{4\pi cn_HU})^{1/2}\;\propto\;L^{1/2}.
\end{equation}
where $N_{ion}$ is the number of ionizing photons. Peterson (1993) showed that 
such a relation may exist for a few Seyfert galaxies with well determined lags. 
The results of Kaspi (1994), including other Seyfert galaxies and PG quasars, are 
not in contradiction with this, but the limited temporal sampling in his data, leaves 
doubtful if this remains valid for the PG quasars, because the high temporal 
frequencies of the variability are quite undersampled. If the lags found here are 
considered to represent a good estimate of the BLR size in 3C390.3, it clearly 
does not fit in the relation defined by these samples. The mean UV luminosity at 
1340\AA~ for 3C390.3 is  L$_\lambda$ = 2~10$^{40}$ erg~s$^{-1}$~\AA$^{-
1}$~ ( h$_0$ = 1.0, as adopted in Peterson,1993), which is the same as for 
NGC5548, but CIV variations for NGC 5548 show a lag of about 8 days versus 
the 63 days for the total line integral for 3C390.3 (see Table 5), suggesting a BLR 
size some 8 times larger in 3C390.3. It should be kept in mind that lags of the 
order of those found in the campaign on NGC 5548 (4-30 days) can not be found 
in the data for 3C390.3 discussed here, since the average sampling interval of 89 
$\pm$ 57 days (ignoring the 4 intervals in excess of one year) will not permit the 
determination of such small delays. This does however not invalidate  the larger 
delays found here, since the sampling for the two large events after 1988 appears 
to be adequate for the characteristic variability time scale of  some 250 days.   

The breakdown of $r\;\propto\;L^{1/2}$ implies that one or more of the 
assumptions, on which the Size--Luminosity relation is based, are invalid for 
3C390.3: (1) The product $n_H$U in 3C390.3 is a factor of 60 less than in other 
AGNs monitored previously; (2) The number of ionizing photons $N_{ion}$ 
striking the BLR in 3C390.3 is 50 times more than simple scaling UV continuum 
would predict.  

More ionizing photons could be a result of the hard ionizing continuum. If the 
continuum incident on the BLR clouds is the same as what we observe -without a 
big blue bump-, the ionizing continuum is much harder than that in other Seyfert 
galaxies as discussed in \S 3. Even if this is the case, the total number of ionizing 
continuum photons in 3C390.3 is estimated to be a only factor $\sim$ 4 larger 
than that in NGC5548 (taking $\alpha_{uvx}$ = 0.8, and 1.5 for 3C390.3 and 
NGC5548 respectively), much less than what is needed to account for the large 
observed delay. With equal covering factors for these two objects, this estimate of 
the ionizing photon flux $N_{ion}$ is consistent with the observed values of 
EW(\la)~ and EW(CIV), which are a factor of $\sim$3-4 larger in 3C390.3 than in 
NGC5548. Alternatively, the continuum in other directions could be much 
stronger than observed, due to the anistropic emission.  However, this would 
cause an increase in the equivalent widths of the emission lines as well. The 
EW(\la) and EW(CIV) limit the degree of anisotropy for the ionizing continuum 
toward BLR. The continuum in the direction of  BLR can not be a factor of 10 
larger for any reasonable value for the covering factor. Therefore, $n_eU$~ must 
be an order of magnitude less in 3C390.3 than in NGC 5548. 

If the continuum emission is isotropic, the total luminosity of 3C390.3 
integrated from optical to X--Ray (assuming it extends to 500 keV) will be 
in the range 0.13 to 1.8 10$^{45}$~erg~s$^{-1}$ (corresponding to the observed 
variability), which is well within the range of the  Eddington luminosity for a 
10$^9$ 
solar mass black hole (0.001-0.01 L$_{Edd}$). The luminosity in the UV and 
optical range is only a fraction of this, and therefore fully compatible with the 
accretion rates which can be sustained in an ion-supported torus (Begelman,1986). 
The infrared spectrum peaking at 25 $\mu$m in this object would than represent  
the synchrotron self absorption of an ion torus, as already suggested by Chen \& 
Halpern (1988). The absence of a thermal big blue bump implies that a possible 
disk is not cooling down via thermal radiation, but rather  via inverse Compton 
scattering of low energy photons, consistent with ion-torus model (Rees et 
al.,1982). 

\subsection{The CIV/\la~ Ratio}
The results shown in Figure 4 for the CIV/\la~ variations can be easily understood 
in the context of the continuum variation as described above in absence of  a big 
blue bump component in the continuum incident on the BLR clouds. In 
NGC5548, F9 and 3C120, both the ratio CIV/\la~ and the EW(CIV) decreases as 
the continuum flux increases (Pogge \& Peterson,1992; Clavel, Wamsteker \& 
Glass,1989; Gondhalekar,1992). This behavior is exactly the opposite of the usual 
prediction and has been interpreted as an indication that the big blue bump does 
not continue to rise with photon energy, but rolls over at energies higher than 10 
eV, and a second hard component is less variable (Binnette et al.,1989; Clavel \& 
Santos-Lle\'o,1990; also Maoz, Peterson \& Netzer,1994). However, Shields et al. 
(1995) showed that such decrease in the CIV/\la~ratio could also result from a 
partially optically thin BLR (their   "Wamsteker--Colina" effect). The results 
found here suggest that we are seeing the effect of the hard continuum on 
a  mixed medium of thin and thick clouds, as suggested by Shields et al. (1995). 
The very steep increase at low luminosity would then correspond to the an 
increase in the ionizing fraction of CIV in the thin cloud, while this will not be 
accompanied by a major increase in the \la~ emission. However, when the 
ionization parameter becomes higher, the thick clouds will start to dominate the 
emission, with the consequent flattening of the   CIV/\la~ ratio, since the intensity 
increase in the thick BLR clouds is similar for both \la~ and CIV. This could e.g. 
be associated with deeper penetration of the \la~ ionization front as compared to 
the CIV ionization front at low brightness levels. The consequence of this scenario 
for 3C390.3 is that, in comparison with most of the previously studied radio--quiet 
AGN's, the optically thin material in the BLR must be of much greater importance 
here. This is the first time that a real difference in  the nature of the BLR clouds in 
a radio--loud and radio--quiet AGN's is indicated.

\subsection{Kinematics of BLR}

With a central source of ionizing radiation, the obvious consequence of the fact 
that the variation in red side of CIV precedes that in the blue side is,  that the gas 
moving away from us is located between the observer and  the source of ionizing 
radiation, while the blue side of the line is originating on the far side of the central 
source, i.e.  gas infall towards the center. A simple spherical symmetrical inflow 
model predicts, $\tau_{blue}\; = \; 3\tau_{red}$. This might still be in agreement 
with the observations with $\tau_{red}\; = \; 57\pm18$~days  and  $\tau_{blue}\; 
= \;110\pm17$~days, but the results suggest more something like $\tau_{blue}\; 
\simeq \; 2\times \tau_{red}$. 
Since many previous observational results do suggest the presence of anisotropy 
in either the radiation field and/or the material distribution, we will for the 
following discussion assume a disk configuration of the material in the BLR. In 
this case {\it some form of rotational symmetry} is indicated and the difference in 
the delays then places the broad-line emitting material on opposite sides of such 
disk. The difference in time delay between the blue and red wings of the lines 
allows us to determine their projection on the sky, which we find to be 
18$^{+10}_{-23}$ degrees (i.e. the angle between the l.o.s. and the direction 
connecting the blue and red wing emitting material, not necessarily equal to the 
inclination of a possible disk, is 72$^{+13}_{-10}$  degrees). The actual size of 
the measured delay then places the BLR material at a distance of  83$\pm$25~ 
lightdays from the central source. Assuming that these regions are in rotation 
around the center of mass associated with the continuum source, the difference in 
velocity between the blue and the red sides is the combined effect of rotational 
and radial motion. Taking the peak in the variance spectrum of the CIV 
line 
to represent the mean line--of--sight velocity of 3000 km~s$^{-1}$ for both  the 
redshifted and blue-shifted material in the BLR, we find a purely rotational  
component of the velocity field of the BLR of v$_{circ}$~= 2850 $^{+100}_{-
700}$ km~s$^{-1}$, combined with an infall velocity of v$_{infall}$~=950 
$^{+1000}_{-500}$ km~s$^{-1}$ (errors represent the uncertainty in the 
projection angle above). Such  a systematic velocity field and infall in the BLR 
also supplies a mechanism to allow the natural removal of angular momentum 
from the BLR as a consequence of the preferential escape from the system of 
material, with the largest random velocities, in the direction of the systematic 
rotation.The existence of ordered rotational motion in bound orbits at a distance of 
R$_{BLR}$= $2.2~10^{17}$ cm from the central source permits, under the 
assumption that we are seeing the disk edge-on, the determination of a 
$\it{lower~limit}$ to the central mass of  M~$_{CM} > 1.3~10^8$~$M\odot$. 
In the presence of ordered bound rotational motion around a central mass the 
FWHM of the broad lines corresponds to the escape velocity from the central 
mass, which places an $\it{upper~limit}$ to the central mass, within the radius of 
the BLR of R$_{BLR}$= 0.07 pc~, of  M~$_{CM} < 4.0~10^8$~$M\odot$.

In conclusion we find here that the central mass is confined between 
$1.3~10^8$~$M\odot~<$ M$_{CM}$ < $4.0~10^8$~$M\odot$, which places the 
BLR in 3C390.3 at a distance of  500$\pm$150~$~R_{Schwarzschild}~<$ 
R$_{BLR}$< 1,500$\pm$500~$R_{Schwarzschild}$, with an orbital period of 
some 150 years, which is only twice the infall time towards the center at the infall 
velocity of 950 km~s$^{-1}$. These limits on the BLR size can be compared with 
those derived by Eracleous and Halpern (1994) on the basis of detailed accretion 
disk model fits to the outer wings of \ha~. They found 
380~$R_{Schwarzschild}~<$ R$_{BLR}$< 1,300~$R_{Schwarzschild}$, with a 
disk inclination (the angle between the line--of--sight and the normal to the disk) 
of 26$^{+4}_{-2}$~degrees. Using the superluminal motions in the jet of 
3C390.3, Eracleous et al. (1996) determined the angle of ejection of the radio 
blobs and the line--of--sight,  
which was found to be between 19 and 33 degrees. Our implicit assumptions for 
the results from the UV data, only required rotational symmetry. However, since 
generally speaking, the broad emission lines should all come from the BLR (even 
though this may be stratified), the only way in which these results are consistent 
with those from the UV lines discussed here, is if the angular projection angle for 
the blue and read parts of the UV lines of 18$^{+10}_{-23}$ degrees is 
considered to be an independent measure of the inclination of the disk required for 
the \ha~ line.
Combing these angular results, we see that in 3C390.3 we have for the first time 
strong {\it observational} evidence that the BLR emission is associated with an 
accretion disk and the superluminal motions take place orthogonal to the plane of 
this disk (98$\pm$12~degrees). In such configuration (see Figure 6), if the full 
line width of Fe $K\alpha$ emission would be due to Doppler broadening from 
the motion in the disk (Eracleous et al.,1996), the Fe $K\alpha$ would be emitted 
in a region at a distance $<400 ~R_{Schwarzschild}$, corresponding to the inner 
edge of the BLR as found from the UV lines. One caveat for this conclusion could 
be the strong evidence for optically thin gas found here (see Figure 4), which may 
have an effect on the modeling of the Fe $K\alpha$ emission, which has not been 
taken into consideration by Eracleous et al. (1996).

The variations in the relative contributions to the total line profile of the 
differently behaved velocity components in 3C390.3 could easily explain 
the large blue shifts in CIV and \la~, often seen in luminous quasars, relative to 
low ionization lines such as OI and MgII (Gaskell,1982; Espey et al.,1989; 
Corbin,1990). Ferland et al.(1992) have shown that the line emission from thick 
clouds is anisotropic with preferred emission from the clouds in the direction of 
the ionizing source (see also Ferland \& Netzer,1979). Our results then 
suggest that the material farthest from us would dominate the line profiles, since 
they are illuminated by the central source from a direction close to our viewing 
angle, giving rise to the strong line strength asymmetries seen in Figure 2 for 
3C390.3.

\section{SUMMARY AND CONCLUSIONS}

We have described the results of an analysis IUE, optical and X--Ray observations 
of 3C390.3. The main results are: \\[2mm]
 
1. The variations of the total intensity of the CIV and \la~lines lag behind those of 
UV continuum by  some 66 days, which is a factor of 8 larger than one would 
expect by simply scaling $r\;\propto\;L^{1/2}$ from the data of other Seyfert 
galaxies. 

2. The red side of both  the \la~ and the CIV lines responds to continuum 
variations faster than the blue side by about 46$\pm$17 days. This results under 
rotational symmetry to a distance of the BLR to the central source of R$_{BLR}$ = 
83$\pm$25  lightdays. Assuming rotational motion, and since the differential 
delay indicates infall, this suggests a large infalling component of 950 
$^{+1000}_{-500}$ km~s$^{-1}$ in the BLR velocity field with a pure 
rotational component of 2850 $^{+100}_{-700}$ km~s$^{-1}$. From this we 
conclude the central mass to be confined to $1.3~10^8 M\odot < M_{CM}< 
4.0~10^8 M\odot$.    

3. From an analysis of the optical, UV and X--Ray spectrum, no big blue bump or 
soft X--Ray excess is found in 3C390.3. The continuum spectrum from the optical 
to the X-rays can be described by a single power law with index $\alpha$  
$\simeq$~ -0.90. 

4. The CIV/\la~ ratio increases as the continuum  brightens and flattens off at high 
continuum levels, consistent with the expected behavior under  photoionization of 
a mixed cloud composition of optically thin and thick clouds, with a constant 
continuum shape. This is also consistent with the absence of a big blue bump in 
the spectral energy distribution of 3C390.3 and indicates for the first time a 
difference in the BLR of a radio--loud AGN as compared to the radio--quiet 
Seyfert I galaxies.

5. The blue part in the CIV, \la~ and Balmer lines originates in the same emitting 
material, but the red peak in \hb~ does not respond to the continuum variations 
and therefore is different, most likely at considerably larger distance.

6. The physical dissociation and the different delays for the blue and red sides of 
the broad emission lines under rotational symmetry, suggests that we see the blue 
part of the BLR clouds  face--on with the red part more in a back--illuminated 
geometry. This supplies a logical explanation for the blue shifts seen in the high 
luminosity QSO's with respect to the narrow lines, since the blue part of the 
originates in clouds seen from the directly illuminated side.

7. Comparing our results with recent ASCA observations by Eracleous et al. 
(1996) our results are fully consistent with theirs if the UV lines are emitted in an 
accretion disk, where the Fe $K\alpha$ is emitted at the inner edge of the disk 
with a disk inclination between 18 and 30 degrees, and the superluminal radio 
blobs moving orthogonal to the plane of the disk associated with the BLR. 
\\[3mm]

\section*{Acknowledgments}
We are indebted to D. Osterbrock, J. Miller and S. 
Veilleux for providing the optical spectra of 3C390.3. TW would like to thank 
Jean Clavel for many stimulating discussions and help in the IUE data processing. 
We thank the anonymous referee for his comments, which significantly 
contributed to the presentation of the data. TW wishes to acknowledge the support 
of the ESA IUE Observatory at VILSPA, and financial support from the Pedan 
Project of Chinese Scientific Committee and Chinese National Natural Science 
Foundation.

\begin{table*}
\centering
\begin{minipage}{110mm}
\caption{IUE Observation Log and Continuum Flux}
\begin{tabular}{rrccc}\hline\hline
No. & Julian date & date & F$_{\lambda}$(1340\AA) & F$_{\lambda}$(1820\AA) \\ 
  & -2,440,000 &    &      10$^{-14}$erg~s$^{-1}$cm$^{-2}$\AA$^{-1}$ &  \\
1 & 3,833& 21Nov78  & 0.65$\pm$0.28 & 0.43$\pm$0.15\\
2 & 3,840& 28Nov78  & 0.89$\pm$0.66 &  $-----$ \\
3 & 3,920& 16Feb79  & 0.49$\pm$0.37 & 0.32$\pm$0.13\\
4 & 3,965& 02Apr79  & 0.40$\pm$0.47 & 0.43$\pm$0.26\\
5 & 3,969& 06Apr79 & 0.59$\pm$0.37 & 0.37$\pm$0.22\\
6 & 4,385& 26May80 & 0.35$\pm$0.33 & 0.34$\pm$0.19\\
7 & 4,577& 04Dec80 & 0.39$\pm$0.33 & 0.37$\pm$0.17\\
8 & 4,600& 27Dec80 & 0.62$\pm$0.24 & 0.45$\pm$0.10\\
9 & 4,634& 30Jan81 & 0.40$\pm$0.35 & 0.37$\pm$0.22\\
10& 5,176& 26Jul82 & 0.31$\pm$0.40 & 0.28$\pm$0.17\\
11& 5,280& 07Nov82 & 0.15$\pm$0.25 & 0.09$\pm$0.42\\
12& 5,421& 28Mar83 & 0.28$\pm$0.26 & 0.24$\pm$0.12\\
13& 5,600& 23Sep83 & 0.51$\pm$0.28 & 0.38$\pm$0.15\\
14& 5,789& 30Mar84 & 0.15$\pm$0.42 & 0.19$\pm$0.32\\
15& 5,856& 05Jun84 & 0.32$\pm$0.24 & 0.18$\pm$0.09\\
16& 5,939& 27Aug84 & 0.35$\pm$0.32 & 0.19$\pm$0.13\\
17& 6,150& 26Mar85 & 0.92$\pm$0.23 & 0.55$\pm$0.11\\
18& 6,672& 30Aug86 & 0.52$\pm$0.39 & 0.48$\pm$0.13\\
19& 6,823& 28Jan87 & 0.43$\pm$0.37 & 0.34$\pm$0.14\\
20& 6,951& 05Jun87 & 0.51$\pm$0.29 & 0.34$\pm$0.13\\
21& 7,019& 12Aug87 & 0.68$\pm$0.33 & 0.53$\pm$0.14\\
22& 7,160& 31Dec87 & 0.54$\pm$0.34 & 0.56$\pm$0.19\\
23& 7,207& 16Feb88 & 0.32$\pm$0.16 & 0.31$\pm$0.18\\
24& 7,607& 22Mar89 & 0.73$\pm$0.28 & 0.41$\pm$0.14\\
25& 7,655& 09May89 & 0.56$\pm$0.25 & 0.36$\pm$0.10\\
26& 7,752& 14Aug89 & 0.71$\pm$0.41 & 0.80$\pm$0.15\\
27& 7,843& 13Nov89 & 0.81$\pm$0.32 & 0.99$\pm$0.19\\
28& 7,920& 29Jan90 & 0.83$\pm$0.28 & 0.55$\pm$0.12\\
29& 7,939& 17Feb90 & 0.78$\pm$0.26 & 0.46$\pm$0.12\\
30& 7,998& 17Apr90 & 0.84$\pm$0.51 & 0.61$\pm$0.20\\
31& 8,083& 11Jul90 & 0.62$\pm$0.17 & 0.39$\pm$0.07\\
32& 8,133& 30Aug90 & 0.49$\pm$0.20 & 0.40$\pm$0.13\\
33& 8,135& 01Sep90 & 0.51$\pm$0.18 & 0.36$\pm$0.09\\
34& 8,253& 28Dec90 & 1.26$\pm$0.31 & 0.86$\pm$0.24\\
35& 8,369& 23Apr91 & 0.99$\pm$0.23 & 0.73$\pm$0.12\\
36& 8,499& 31Aug91 & 2.37$\pm$0.31 & 1.35$\pm$0.12\\
37& 8,634& 13Jan92 & 1.43$\pm$0.26 & 0.89$\pm$0.15\\
38& 8,661& 09Feb92 & 0.96$\pm$0.23 & 0.67$\pm$0.09\\
39& 8,762& 20May92 & 0.62$\pm$0.24 & 0.50$\pm$0.14\\ \hline
\end{tabular}
\end{minipage}
\end{table*}

\begin{table*}
 \centering
 \begin{minipage}{110mm}
\caption {UV Line Fluxes}
\begin{tabular}{lcccccccc}\hline\hline
No. & & &Ly$\alpha$& & & & CIV & \\ 
    & blue & centre & red & total & blue & centre & red & total \\ \hline
\multicolumn{8}{c}{        10$^{-14}$~erg~cm$^{-2}$~s$^{-1}$ } \\
  1&  80$\pm$5 & 172$\pm$9 &  60$\pm$5 & 348$\pm$21 &  38$\pm$4 &  65$\pm$4 &  13$\pm$4 & 131$\pm$17 \\
  2&  47$\pm$5 & 147$\pm$8 &  80$\pm$6 & 299$\pm$53 &  57$\pm$15 &  73$\pm$10 &  46$\pm$15 & 207$\pm$66 \\
  3&  62$\pm$4 & 155$\pm$8 &  67$\pm$5 & 306$\pm$21 &  38$\pm$4 &  67$\pm$4 &  34$\pm$4 & 157$\pm$20 \\
  4&  67$\pm$7 & 173$\pm$9 &  66$\pm$7 & 336$\pm$27 &  45$\pm$6 &  60$\pm$5 &  36$\pm$6 & 190$\pm$28 \\
  5&  61$\pm$9 & 140$\pm$9 &  54$\pm$9 & 272$\pm$22 &  45$\pm$5 &  59$\pm$4 &  17$\pm$5 & 140$\pm$23 \\
  6&  57$\pm$5 & 145$\pm$8 &  48$\pm$4 & 274$\pm$20 &  13$\pm$4 &  42$\pm$3 &   8$\pm$4 &  71$\pm$19 \\
  7&  46$\pm$5 & 132$\pm$7 &  53$\pm$5 & 258$\pm$19 &  26$\pm$4 &  51$\pm$4 &  19$\pm$4 & 111$\pm$19 \\
  8&  44$\pm$4 & 123$\pm$6 &  66$\pm$4 & 251$\pm$16 &  22$\pm$3 &  47$\pm$3 &  16$\pm$3 & 103$\pm$13 \\
  9&  51$\pm$4 & 120$\pm$6 &  46$\pm$4 & 229$\pm$20 &  34$\pm$5 &  57$\pm$4 &  29$\pm$5 & 144$\pm$22 \\
 10&  45$\pm$5 & 142$\pm$8 &  49$\pm$5 & 254$\pm$21 &  15$\pm$5 &  46$\pm$4 &  10$\pm$5 &  74$\pm$21 \\
 11&  35$\pm$3 & 113$\pm$6 &  34$\pm$3 & 192$\pm$21 &  14$\pm$5 &  39$\pm$4 &  11$\pm$5 &  75$\pm$25 \\
 12&  45$\pm$3 & 111$\pm$6 &  33$\pm$3 & 201$\pm$15 &  13$\pm$3 &  35$\pm$3 &  13$\pm$3 &  69$\pm$14 \\
 13&  49$\pm$4 & 121$\pm$6 &  44$\pm$4 & 226$\pm$17 &  29$\pm$4 &  40$\pm$3 &  20$\pm$4 & 103$\pm$16 \\
 14&  40$\pm$6 & 110$\pm$6 &  33$\pm$5 & 193$\pm$23 &  17$\pm$6 &  39$\pm$4 &   9$\pm$6 &  83$\pm$27 \\
 15&  34$\pm$3 &  90$\pm$5 &  25$\pm$3 & 151$\pm$12 &  11$\pm$3 &  30$\pm$2 &  11$\pm$3 &  61$\pm$12 \\
 16&  16$\pm$3 &  70$\pm$4 &  24$\pm$3 & 113$\pm$14 &   9$\pm$4 &  20$\pm$3 &   2$\pm$4 &  32$\pm$16 \\
 17&  53$\pm$5 & 109$\pm$6 &  46$\pm$4 & 239$\pm$15 &  37$\pm$3 &  42$\pm$3 &  32$\pm$3 & 144$\pm$14 \\
 18&  44$\pm$4 & 105$\pm$6 &  44$\pm$4 & 212$\pm$18 &  22$\pm$4 &  29$\pm$3 &  18$\pm$4 &  90$\pm$20 \\
 19&  67$\pm$5 & 121$\pm$7 &  46$\pm$5 & 248$\pm$19 &  40$\pm$5 &  41$\pm$3 &  29$\pm$4 & 139$\pm$20 \\
 20&  73$\pm$5 & 125$\pm$7 &  44$\pm$3 & 257$\pm$18 &  25$\pm$4 &  39$\pm$3 &  23$\pm$4 & 104$\pm$16 \\
 21&  59$\pm$4 & 126$\pm$7 &  43$\pm$4 & 237$\pm$18 &  36$\pm$4 &  47$\pm$3 &  29$\pm$4 & 129$\pm$18 \\
 22&  55$\pm$5 & 110$\pm$6 &  43$\pm$4 & 207$\pm$18 &  17$\pm$4 &  45$\pm$4 &  16$\pm$4 &  80$\pm$19 \\
 23&  69$\pm$4 & 108$\pm$6 &  42$\pm$3 & 235$\pm$15 &  31$\pm$3 &  40$\pm$3 &  18$\pm$3 &  93$\pm$13 \\
 24& 102$\pm$6 & 149$\pm$8 &  66$\pm$5 & 334$\pm$21 &  75$\pm$5 &  70$\pm$4 &  40$\pm$4 & 209$\pm$19 \\
 25&  99$\pm$5 & 146$\pm$7 &  61$\pm$4 & 326$\pm$19 &  68$\pm$4 &  62$\pm$4 &  33$\pm$3 & 179$\pm$15 \\
 26&  96$\pm$7 & 154$\pm$8 &  80$\pm$6 & 367$\pm$24 &  64$\pm$5 &  73$\pm$5 &  49$\pm$5 & 215$\pm$23 \\
 27& 112$\pm$6 & 180$\pm$9 &  81$\pm$5 & 411$\pm$25 &  81$\pm$6 &  81$\pm$5 &  50$\pm$5 & 243$\pm$22 \\
 28&  86$\pm$5 & 158$\pm$8 &  92$\pm$5 & 368$\pm$22 &  67$\pm$5 &  75$\pm$4 &  48$\pm$4 & 217$\pm$18 \\
 29& 102$\pm$6 & 147$\pm$8 &  69$\pm$5 & 335$\pm$20 &  80$\pm$5 &  70$\pm$4 &  35$\pm$4 & 206$\pm$17 \\
 30&  82$\pm$5 & 117$\pm$6 &  62$\pm$5 & 285$\pm$25 &  62$\pm$6 &  62$\pm$5 &  36$\pm$6 & 181$\pm$27 \\
 31&  78$\pm$4 & 108$\pm$5 &  53$\pm$3 & 257$\pm$15 &  48$\pm$3 &  51$\pm$3 &  28$\pm$2 & 143$\pm$11 \\
 32&  64$\pm$4 & 123$\pm$6 &  47$\pm$3 & 250$\pm$16 &  35$\pm$3 &  42$\pm$3 &  25$\pm$3 & 117$\pm$13 \\
 33&  65$\pm$4 & 124$\pm$6 &  53$\pm$4 & 257$\pm$15 &  36$\pm$3 &  41$\pm$3 &  23$\pm$2 & 115$\pm$12 \\
 34&  76$\pm$7 & 123$\pm$7 &  48$\pm$6 & 261$\pm$20 &  43$\pm$5 &  61$\pm$4 &  40$\pm$5 & 183$\pm$22 \\
 35&  96$\pm$6 & 141$\pm$7 &  65$\pm$4 & 330$\pm$19 &  56$\pm$4 &  62$\pm$4 &  42$\pm$3 & 191$\pm$16 \\
 36& 104$\pm$7 & 157$\pm$8 &  94$\pm$7 & 389$\pm$23 &  79$\pm$5 &  84$\pm$5 &  69$\pm$5 & 268$\pm$20 \\
 37& 158$\pm$8 & 195$\pm$10 & 101$\pm$6 & 489$\pm$27 & 116$\pm$7 & 104$\pm$6 &  79$\pm$5 & 342$\pm$23 \\
 38& 150$\pm$8 & 182$\pm$9 &  96$\pm$6 & 464$\pm$25 & 107$\pm$6 &  92$\pm$5 &  66$\pm$4 & 307$\pm$19 \\
 39& 122$\pm$7 & 162$\pm$8 &  76$\pm$5 & 382$\pm$22 &  96$\pm$6 &  80$\pm$4 &  49$\pm$4 & 244$\pm$19 \\
\hline
\end{tabular}
\end{minipage}
\end{table*}

\begin{table*}
 \centering
 \begin{minipage}{130mm}
\caption {X-Ray Data and nearest UV Continuum Flux}
\begin{tabular}{llccccc}\hline\hline
Date & Instrument & A(0.01)$^a$ & $\Gamma$ & $\Delta$t$_{uvx}$ & N$_H^b$ & $\chi^2$/d.o.f. \\ \hline
781121 & HEAO2/MPC & 0.37 & 1.3$\pm$0.24 & 0 & & \\
781224 & HEAO1/A2 & 0.74 & 1.65$\pm$0.30 &-34 & & \\
781224 & HEAO1/A4 & 0.78 & 1.40$\pm$0.37 & -34 & &  \\
800101 & HEAO2/IPC &0.60$^{+0.77}_{0.26}$ & 2.75$^{+0.53}_{-0.44}$ & 60.3$\pm$1.9  &  \\
800408 & HEAO2/MPC & 0.18 &1.40$^{+0.51}_{-0.37}$ & 56 &  & \\
830904 & EXOSAT/ME & 0.69$\pm$0.23 & 1.68$^{+0.20}_{-0.08}$ &19 & & 36/33  \\
840602 & EXOSAT/ME & 0.29$^{+0.33}_{-0.14}$ & 2.5$\pm$0.7 & 2 & & 27/18 \\
850202 & EXOSAT/ME & 0.51$\pm$0.15 &1.75$\pm$0.2 & 55 & & 37/35\\
851107 & EXOSAT/ME & 0.57$\pm$0.12 & 1.70$\pm$0.2 & -- & & 26/32 \\
860316 & EXOSAT/ME & 0.44$\pm$0.01 & 1.60$\pm$0.10 & 168 & & 27/32 \\
860317 & EXOSAT/ME & 0.43$\pm$0.05 & 1.57$\pm$0.08 & 167 & & 21/36  \\
881111 & Ginga & 1.29$\pm$0.03 & 1.77$\pm$0.01 & 120 & & 40.1/37\\
910214 & Ginga & 0.51$\pm$0.02 & 1.54$\pm$0.02 & 47 & & 45.0/35  \\
910330 & ROSAT & 0.58$\pm$0.01 & 1.86$\pm$0.07 & 24 & 5.93$\pm$0.35 & 27.1/17\\
920411 & ROSAT & 1.00$\pm$0.06 & 1.88$\pm$0.08 & 38 & 5.84$\pm$0.85 & 13.0/15\\ 
930411 & ROSAT & 0.39$\pm$0.06 & 1.90$\pm$0.09 & -- & 6.74$\pm$0.70 & 18.4/14\\
930412 & ROSAT & 0.41$\pm$0.03 & 2.21$\pm$0.11 & -- & 8.14$\pm$1.35 & 7.75/13\\ \hline
\end{tabular}

a. A(0.01) = A/(0.01 photon~cm$^{-2}$s$^{-1}$keV$^{-1}$) at 1 keV\\
b. in unit of 10$^{20}$~cm$^{-2}$\\	

\end{minipage}
\end{table*}

\begin{table*}
 \centering
 \begin{minipage}{100mm}
\caption{Variability Parameters}
\begin{tabular}{lccccc}\hline\hline
Feature & Mean flux & F$_{var}$ & R$_{max}$ & R$_{max}(>JD7600)$ \\ \hline
F$\lambda$(1350\AA) & 0.91  &  0.42$\pm$0.09 & 15.8 & 7.4 \\
F$\lambda$(1800\AA) & 0.38 & 0.38$\pm$0.05 & 7.1 & 4.4 \\
\la~ total & 344 & 0.20$\pm$0.02 & 4.3 & 2.4 \\
\la~ blue &  100 & 0.26$\pm$0.02 & 9.9 & 2.9 \\
\la~ core & 148 & 0.16$\pm$0.01 & 2.8 & 1.9 \\
\la~ red & 72 & 0.24$\pm$0.02 & 4.2 & 2.9\\
CIV total & 210 & 0.28$\pm$0.02 & 11.7 & 4.3 \\
CIV blue & 70 & 0.33$\pm$0.02 & 12.9 & 6.8 \\
CIV core & 69 & 0.24$\pm$0.02 & 5.2 & 2.7 \\
CIV red & 45 & 0.35$\pm$ 0.03 & $::$ & 4.9\\ \hline
\end{tabular}
 \end{minipage}
\end{table*}

\begin{table*}
 \centering
 \begin{minipage}{100mm}
\caption{Main Characteristics of Cross Correlations}
\begin{tabular}{llcc}\hline\hline
  First Series  & Second Series  & $\Delta t_{peak}$ (Days) & $r_{max}$ \\
   &   & (1) &  \\ 
\hline
F$_\lambda$(1800\AA) & \la~ blue & 108$^{+15}_{-21}$ & 0.84\\
F$_\lambda$(1800\AA) & \la~ core & 88$^{+49}_{-37}$ & 0.78\\
F$_\lambda$(1800\AA) & \la~ red & 66$^{+22}_{-21}$ & 0.84\\
F$_\lambda$(1800\AA) & \la~ total & 69$^{+50}_{-19}$ & 0.81\\
F$_\lambda$(1800\AA) & CIV blue & 111$^{+18}_{-13}$ & 0.84\\
F$_\lambda$(1800\AA) & CIV core & 71$^{+49}_{-31}$ & 0.82\\
F$_\lambda$(1800\AA) & CIV red & 49$^{+13}_{-17}$ & 0.87\\
F$_\lambda$(1800\AA) & CIV total & 63$^{+57}_{-20}$ & 0.84\\
\la~ red             & \la~ blue & 48$^{+20}_{-19}$ & 0.94\\
CIV red              & CIV blue  & 45$^{+17}_{-18}$ & 0.85\\
\hline 
\end{tabular}

(1) The formal errors (Gaskell and Peterson, 1987) in the 

correlation functions (CCF and DCF) are $\pm$ 15 days.   

\end{minipage}
\end{table*}

\newpage
\begin{figure*}
\centering
\begin{minipage}{15.1cm}
\psfig{figure=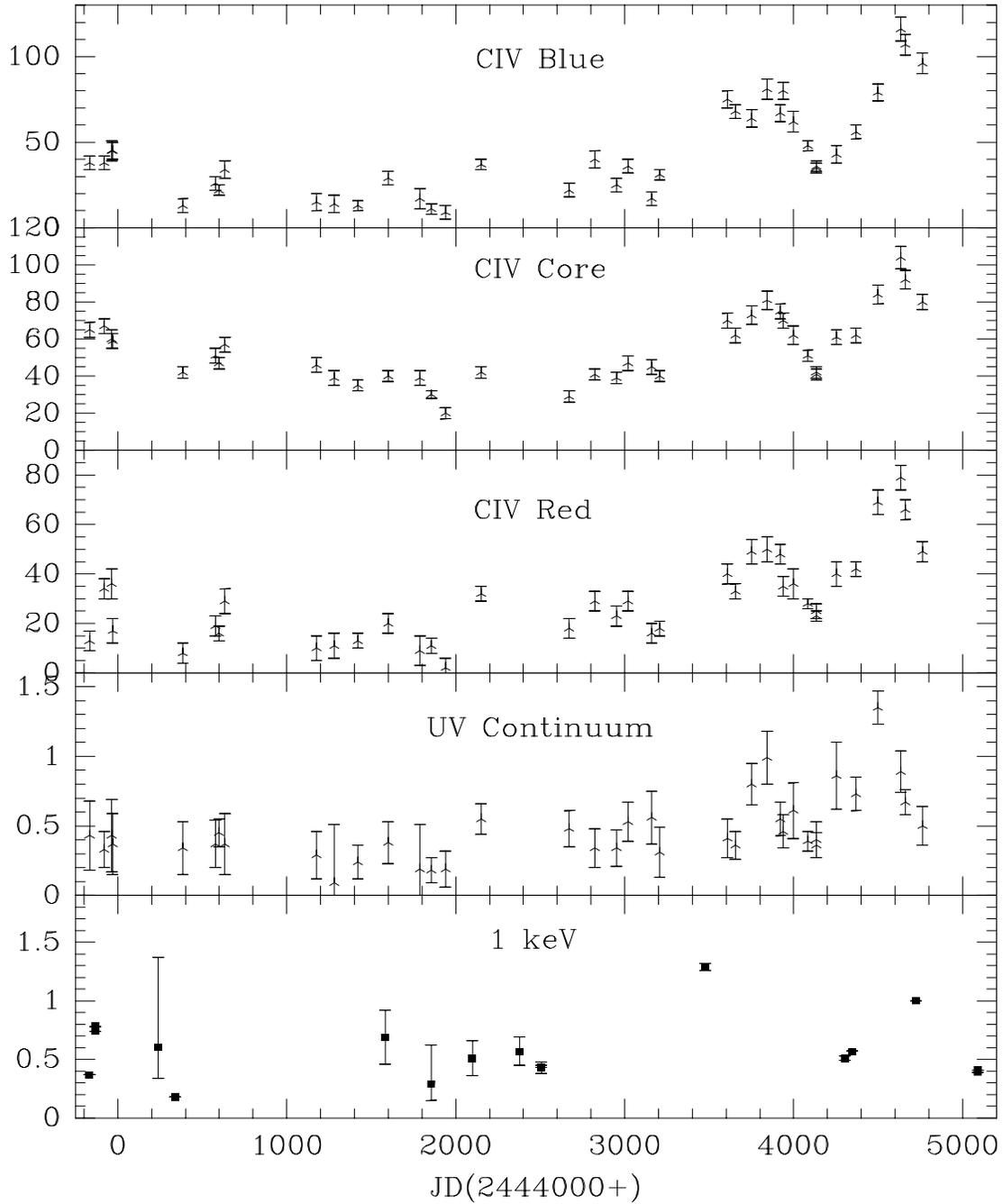,height=19.0cm,width=15.0cm
       ,bbllx=3.0cm,bblly=5.0cm,bburx=18cm,bbury=25cm}
\caption{The lightcurves of 3C390.3 in the X--ray (Table 3),
UV continuum (Table 1) and the CIV line (Table 2). The CIV line is
shown here split up in three velocity domains
(blue: $-7000$~to $-1800$ km~s$^{-1}$; core: $-1800$~to $+1800$ km~s$^{-1}$
and red: $+1800$~to $+7000$ km~s$^{-1}$). The X-ray lightcurve shows the
presence of a large increase in brightness Nov. 1988 (JD6500). At this time
no UV observations were made, but the effects of this event can still
be seen in the blue lightcurve of CIV at (JD6600). The fluxes are in the
rest frame of 3C390.3 and are in units of
10$^{-14}$~erg~s$^{-1}$~cm$^{-2}$~\AA$^{-1}$ for the continuum and
10$^{-14}$~erg~s$^{-1}$~cm$^{-2}$ for the emission lines.}
\end{minipage}
\end{figure*}

\newpage
\begin{figure*}
\centering
\begin{minipage}{15.1cm}
\psfig{figure=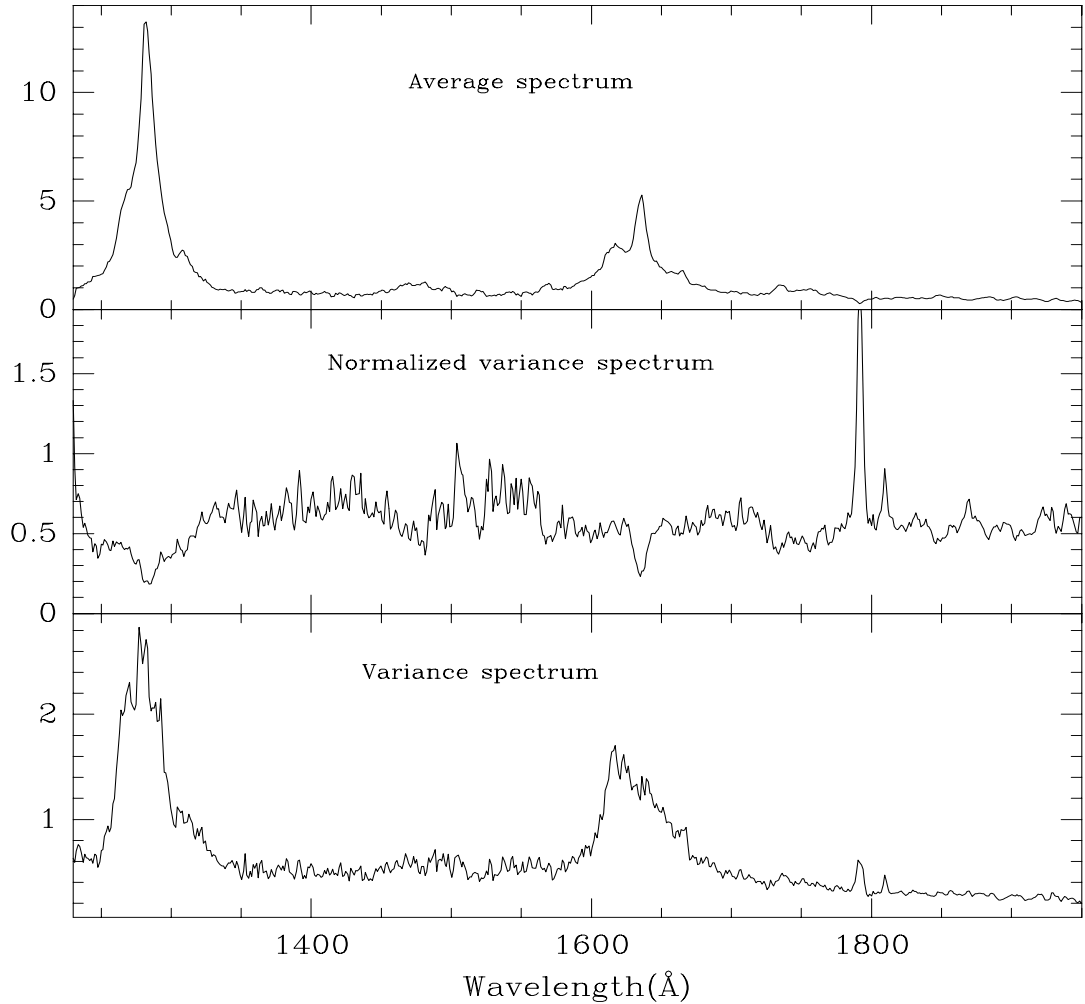,height=15.0cm,width=15.0cm
       ,bbllx=1.0cm,bblly=5.0cm,bburx=20cm,bbury=26cm }
\caption{The top panel shows the average SWP spectrum of 3C390.3 and
the bottom panel shows the variance spectrum (the narrow peaks in the variance
near 1800\AA are due to a reseau mark in the camera). It is clear that
variability exists over the full spectral range. Note the strong assymmetry
in the variance spectrum of especially CIV. The middle panel shows the
normalized variance spectrum (i.e. the quotient of the bottom spectrum with
the top one). The flat--bottomed troughs at the position of \la~and
CIV show that the amplutide of the variabilty is quite similar over the
full width of the broad lines even though the line shape is quite
asymmetric with respect to the systemic velocity.}
\end{minipage}
\end{figure*}

\begin{figure*}
\centering
\begin{minipage}{10.1cm}
\psfig{figure=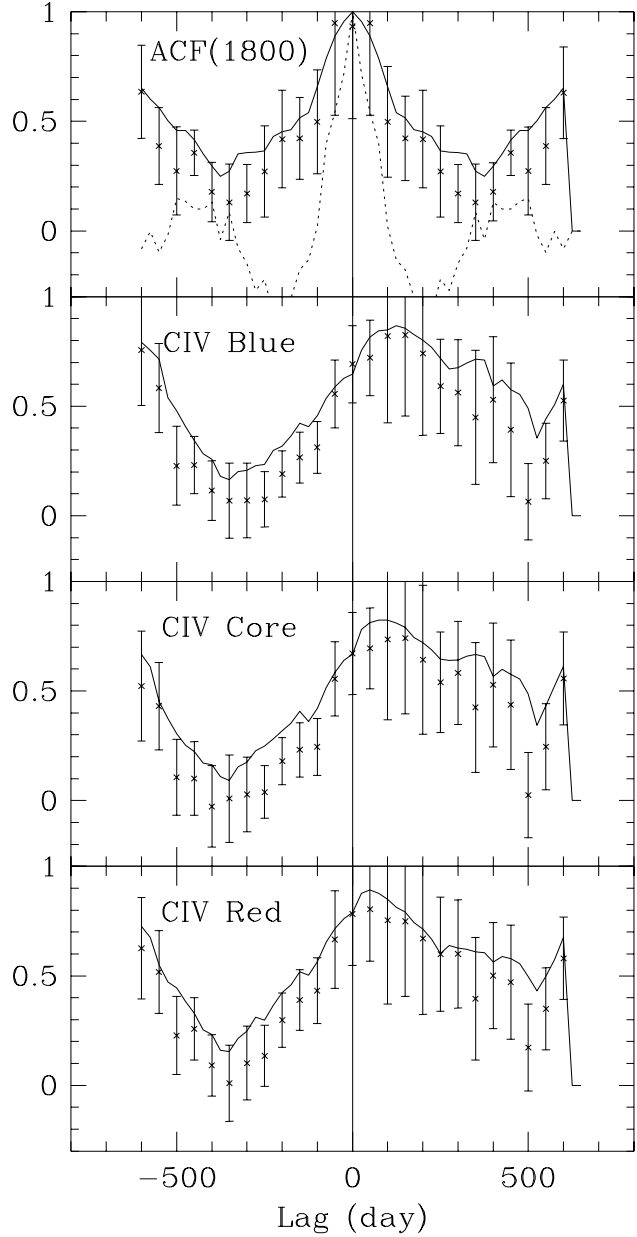,height=18.0cm,width=9.0cm
       ,bbllx=4.0cm,bblly=1.0cm,bburx=17cm,bbury=28cm}
\caption{ From top to bottom this figure shows the ACF at 1800\AA~
with the ACF for the sampling window (top), and the CCF for CIV with the continuum
in the velocity range chosen (see also Figure 1 and text).
All boxes show both the  Interpolated cross-correlation (solid curves)
and the discrete correlation functions (with error bars).
The sampling window autocorrelation function is shown dashed in the top panel.
For reference the
vertical line in the figure shows the position associated with no delay
in the line response. The difference in the delay between the red and blue
sides of the line can be easily discerned and is also present in the
direct CCF between the red and blue sides of the line. The correlation
functions for \la~ are very similar to those of CIV (see also Table 5).}
\end{minipage}
\end{figure*}
\begin{figure*}
\centering
\begin{minipage}{17.0cm}
\psfig{figure=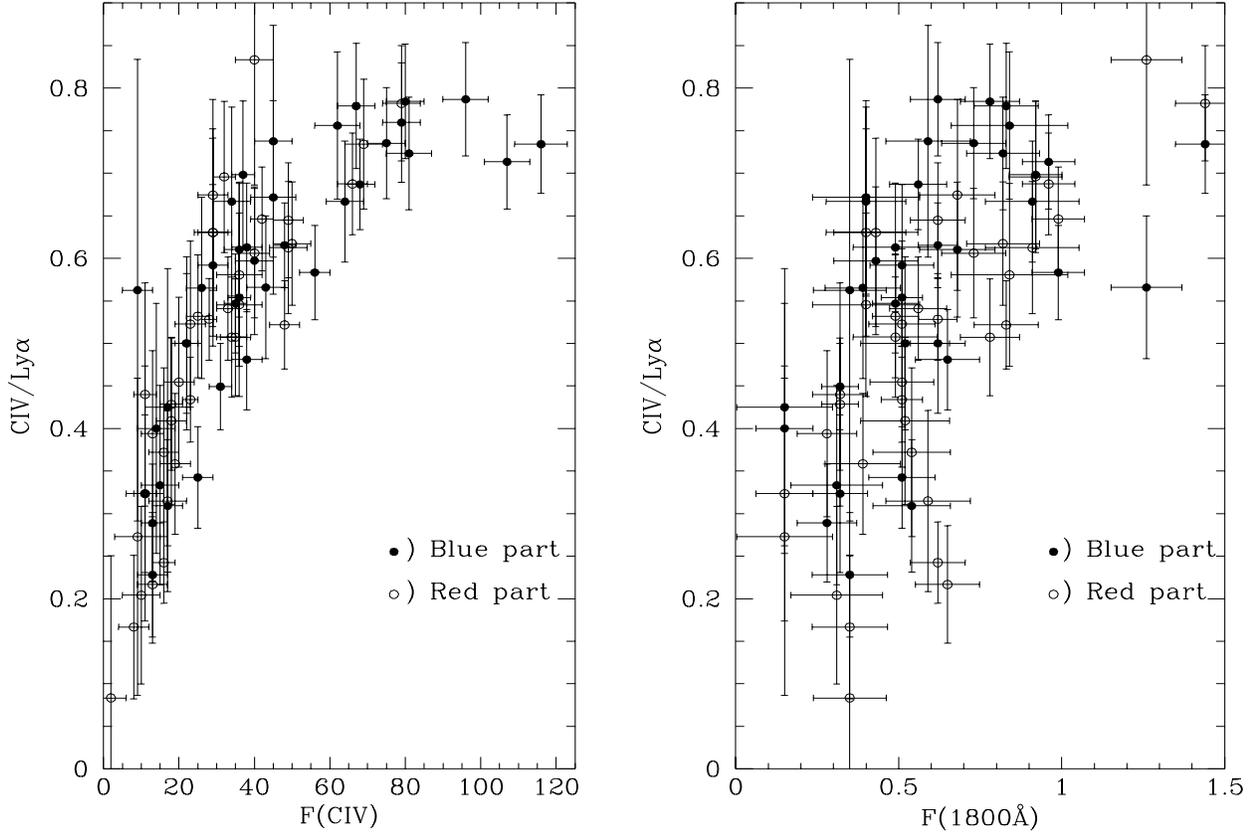,height=12.0cm,width=17.0cm
       ,bbllx=5.0cm,bblly=2.0cm,bburx=21cm,bbury=28cm,angle=270}
\caption{This figure shows on the right--hand side the relation between the
CIV/\la~ ratio and the continuum  and on the left--hand side with the CIV
intensity itself, for both the blue wing (filled symbols) for the red wing
(open symbols). Note that the scatter in the left diagram is much less
that in the continuum diagram on the right. This is predominantly due to the
fact that time delay effects are absent in the left--hand diagram. Note also
that the CIV/\la~ rises steeply with the continuum as measured by
the CIV flux, but appears to remain constant above $I_{CIV} > 60$ (corresponding
to $F_{1800\AA}~ >$ 0.6~10$^{-14}$~erg~s$^{-1}$~cm$^{-2}$~\AA$^{-1}$). This
is a strong indication of a mixed population of optically thin and optically
thick BLR couds. The fluxes are in the rest frame of 3C390.3 and are in
units of 10$^{-14}$~erg~s$^{-1}$~cm$^{-2}$~\AA$^{-1}$ for the continuum
and 10$^{-14}$~erg~s$^{-1}$~cm$^{-2}$ for the emission lines.}
\end{minipage}
\end{figure*}

\begin{figure*}
\centering
\begin{minipage}{10.0cm}
\psfig{figure=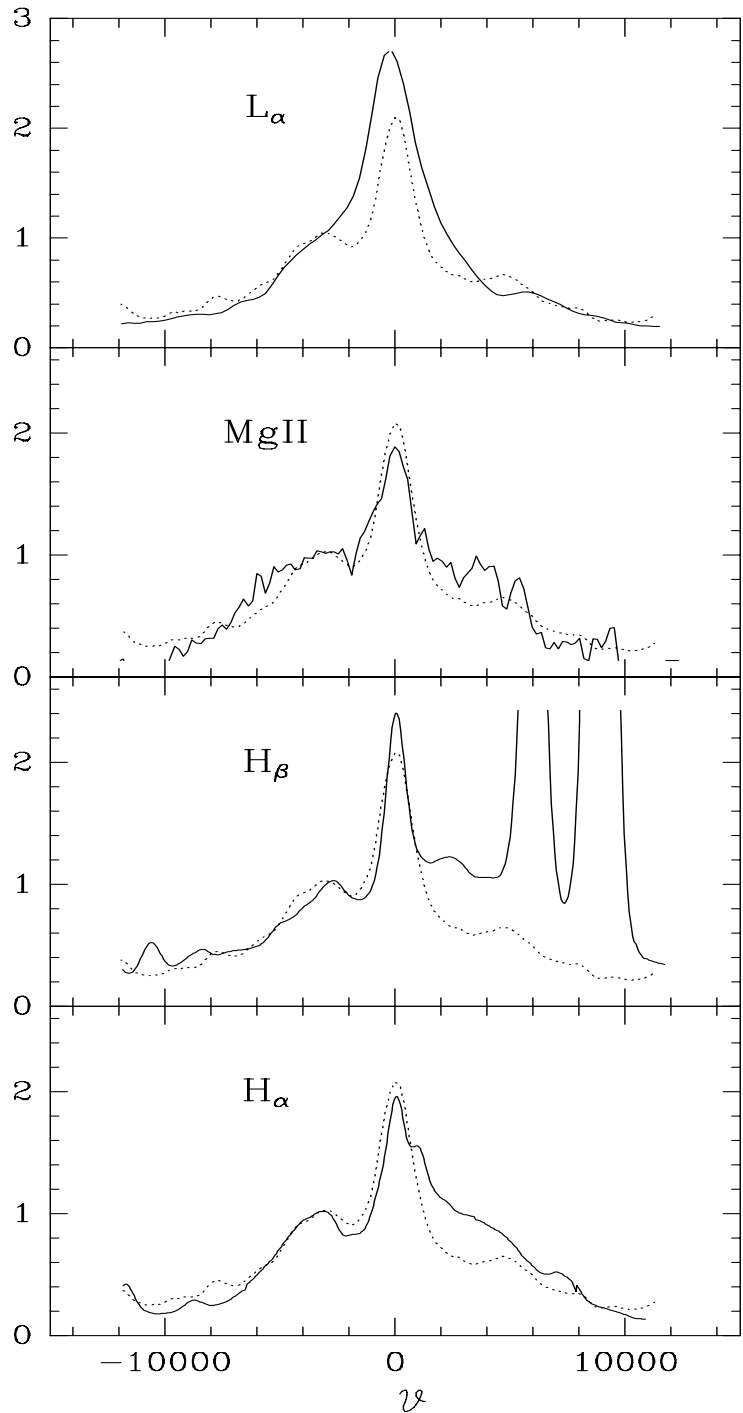,height=20.0cm,width=10.0cm
       ,bbllx=6.0cm,bblly=2.0cm,bburx=15cm,bbury=26cm}
\caption{Average line profiles over the period 1978--1991 for the
UV lines and 1975--1988 for the optical lines. We show the profiles of
\la~,CIV, MgII, \hb~ and \ha~. The CIV line profile is shown as a dotted line.
This diagram clearly illustrates the difference between the Balmer lines and
the UV lines. Both \ha~ and \hb~ show a strong red peak. This is the part of
the Balmer lines which does not respond to the UV continuum variations and is
therefore not associated with the BLR (see also Veilleux and Zheng, 1991).}
\end{minipage}
\end{figure*}

\begin{figure*}
\centering
\begin{minipage}{10.0cm}
\psfig{figure=3C390.3.fig5.ps,height=0.5cm,width=10.0cm
       ,bbllx=1.0cm,bblly=0.0cm,bburx=1.5cm,bbury=0.5cm}
\caption{ {\bf (not included)}
This figure shows schematically the observed configuration of
the BLR of 3C390.3. The angles indicated, are the angles with respect to the
line--of--sight (l.o.s.) determined from the observed superluminal motion and
from the observed delays between the red and blue side of the
UV lines. Since these angles correspond with those derived from the
accretion disk model fit for the wings of \ha~ by Eracleous and Halpern (1994),
and the UV lines suggest a strong rotational, infalling velocity field, these
independent results are only consistent, if the UV lines from the BLR
are associated with an accretion disk as indicated.}
\end{minipage}
\end{figure*}

\end{document}